\documentclass{iopart}
\usepackage{iopams,bm,cite}

\begin{document}

\title{Multicomplementary operators
via finite Fourier transform}

\author{Andrei B~ Klimov$\dag$,
Luis L ~S\'{a}nchez-Soto$^\ddag$
and Hubert de Guise$^\ddag$}

\address{$^\dag$ Departamento de F\'{\i}sica,
Universidad de Guadalajara, Revoluci\'on~1500,
44420~Guadalajara, Jalisco, Mexico}

\address{$\ddag$ Department of Physics,
Lakehead University, Thunder Bay,
Ontario P7B 5E1, Canada}

\begin{abstract}
A complete set of $d+1$ mutually unbiased
bases exists in a Hilbert spaces of dimension
$d$, whenever $d$ is a power of a prime. We
discuss a simple construction of $d+1$ disjoint
classes (each one having $d-1$ commuting
operators) such that the corresponding
eigenstates form sets of unbiased bases.
Such a construction works properly
for prime dimension. We investigate
an alternative construction in which
the real numbers that label the classes
are replaced by a finite field having
$d$ elements. One of these classes is
diagonal, and can be mapped to cyclic
operators by means of the finite Fourier
transform, which allows one to understand
complementarity in a similar way as for
the position-momentum pair in standard
quantum mechanics. The relevant examples
of two and three qubits and two qutrits
are discussed in detail.
\end{abstract}


\eqnobysec

\section{Introduction}

The concept of complementarity is a direct
nontrivial consequence of the superposition
principle and distinguishes purely quantum
systems from those that may be accurately
treated classically. Therefore, a thorough
understanding of this idea is of fundamental
importance for a correct interpretation of
quantum mechanics~\cite{Wheeler83}.

In short, Bohr idea of complementarity could
be loosely formulated by stating that, in
order to understand a quantum phenomenon
completely we need a combination of mutually
exclusive properties: the precise knowledge of
one of them implies that all possible
outcomes in the other are equally probable.

Perhaps the best textbook illustration of
complementarity is that the observation of
interference and the knowledge of the path
followed by the interfering particle are
mutually exclusive. This is often expressed
as the statement that all quantum objects
exhibit particle-like or wave-like behavior
under different experimental conditions.
Standard examples, such as Einstein recoiling
slit~\cite{Cohen90}, Feynman light-scattering
arrangement~\cite{Feynman65} or Heisenberg
microscope~\cite{Heisenberg30}, are always
explained in terms of position and momentum.
It is thus natural that links between canonical
conjugacy and complementarity have been
fully explored~\cite{Tan93,Wiseman97,Luis98}.

However, most if not all of the recent examples
of complementarity involve finite-dimensional
systems (for a complete and up-to-date review
see~\cite{Vourdas04}). Note that, contrary
to what one would expect, the finite
dimensionality of the system space introduces
difficulties when dealing with uncertainty
relations, because the commutators between
complementary observables are operators instead
of $c$-numbers, which makes the analysis
of the problem more involved. The situation
has been recently addressed for two-dimensional
spaces~\cite{Luis01}; however, there are subtle
aspects that cannot be fully encompassed by
these easy-to-understand systems.

In finite-dimensional systems, complementarity
is tantamount to \textit{unbiasedness}~\cite{Schwinger60,Wootters87}:
each eigenstate of any measurement is an
equal-magnitude superposition of the
eigenstates of any of the complementary
measurements. This leads naturally to the
concept of mutually unbiased bases (MUBs),
which have recently been considered with
an increasing interest because of the central
role they play not only in understanding
complementarity~\cite{Kraus87,Lawrence02,Chaturvedi02},
but also in specific quantum information
tasks, such as protocols of quantum
cryptography~\cite{Bechmann00,Cerf02},
Wigner functions in discrete phase
spaces~\cite{Wootters04,Gibbons04},
or the so-called Mean King
problem~\cite{Vaidman87,Englert01,Aravind03,Schulz03,Paz04}.

For a $d$-dimensional system it has been found
that the maximum number of MUBs cannot be
greater than $d +1$ and this limit is reached
if $d$ is prime~\cite{Ivanovic81} or power of
prime~\cite{Wootters89,Calderbank97}. Remarkably
though, there is no known answer for any other
values of $d$, not even for $d = 6$. Recent
works have suggested that the answer to this
question may well be related with the
non-existence of finite projective planes
of certain orders~\cite{Saniga04,Bengtsson04} or with
the problem of mutually orthogonal Latin squares
in combinatorics~\cite{Zauner99,Wootters04b}.

Quite recently, a number of papers have addressed
the explicit construction of MUBs for
dimensions that are prime or composite (i. e.,
power of a prime), exploiting different algebraic
properties~\cite{Ban02,Pitt03,Kla03,Lawrence04,Durt04,Wojcan04}.
In this paper we give an explicit construction
with a different method that resorts to
elementary notions of finite field theory.
We wish to emphasize the distinct features
of our approach: first, we recall
that complementarity for the position-momentum
pair is implemented by the Fourier transform,
which exchanges both operators. Therefore,
we construct classes of maximally commuting
operators and map them using the finite
Fourier transform and an additional
diagonal operator. In consequence, we
obtain in a systematic way the whole family
of complementary operators and not merely MUBs.
Additionally, the final expression for these
MUBs is compact and can be immediately expressed
in different bases, in some of which they
appear as tensor products of generalized
Pauli matrices. In summary, we hope that
our unified construction provides a simple
picture of complementarity for both prime
and composite dimensions.

\section{Complementary operators
in prime dimension}

We consider a system living in a Hilbert
space $\mathcal{H}_{d}$, whose dimension $d$ is
a prime number. It is useful to choose a
computational basis $| n \rangle $ (where $n  =
0, \ldots , d-1$) in $\mathcal{H}_{d}$ and
introduce the basic operators
\begin{eqnarray}
\label{CC}
X | n \rangle & = & |n + 1 \rangle ,
\nonumber  \\
&& \\
Z | n \rangle & = & \omega^{n} | n \rangle ,
\nonumber
\end{eqnarray}
where
\begin{equation}
\omega =\exp (2\pi i/d)
\end{equation}
is a $d$th root of the unity and addition and
multiplication must be understood modulo $d$.
These operators $X$ and $Z$, which are
generalizations of the Pauli matrices, were
studied by Patera and Zassenhaus~\cite{Pat88}
in connection with additive quantum numbers,
and have been used recently by many authors in
a variety of applications~\cite{Got01,Bar02,Gal88}.
They generate a group under multiplication known
as the generalized Pauli group and obey
\begin{equation}
Z X = \omega X Z ,
\end{equation}
which is the finite-dimensional version of the
Weyl form of the commutation relations.

According to the ideas in \cite{Ban02},
we can find $d+1$ disjoint classes (each one
having $d-1$ commuting operators) such that
the corresponding eigenstates form sets of
MUBs. The explicit construction starts with
the following sets of operators:
\begin{eqnarray}
\label{SCop}
\{ Z^{k} \} ,
\qquad \qquad
k = 1,\ldots ,d-1, \nonumber \\
& & \\
\{ (XZ^{m})^{k} \} ,
\qquad k = 1, \ldots , d-1,
\quad m = 0, \ldots , d-1 .
\nonumber
\end{eqnarray}
One can easily check that
\begin{eqnarray}
\label{pairwise}
\Tr ( Z^k Z^{k^\prime}{}^\dagger ) =
d \, \delta_{k k^\prime} ,
\qquad
\Tr ( X^k X^{k^\prime} {}^\dagger ) =
d \, \delta_{k k^\prime}, \nonumber  \\
& & \\
\Tr [ (X Z^m)^k (X Z^{m^\prime})^{k^\prime}
{}^\dagger ] = d \, \delta_{k k^\prime}
\delta_{m m^\prime} .
\nonumber
\end{eqnarray}
These pairwise orthogonality relations
indicate that, for every value of $m$,
we generate a maximal set of $d-1$ commuting
operators and that all these classes are disjoint.
In addition, the common eigenstates of each
class $m$ form different sets of unbiased bases.
We shall refer to these classes as multicomplementary.

We would now like to make the very important
observation that, starting from $Z$, it is
possible to obtain any element of the form
$(X Z^{m})^{k}$ by using a combination of
only two operators $F$ and $V$ defined as
follows: $F$ is the finite Fourier
transform \cite{Ip02}
\begin{equation}
\label{FT1}
F = \frac{1}{\sqrt{d}}
\sum_{n,n^\prime = 0}^{d-1}
\omega^{n n^\prime} \,
|n \rangle \langle n^{\prime }|,
\end{equation}
and $V$ is the diagonal transformation
(assuming $d$ is odd)
\begin{equation}
\label{DT1}
V = \sum_{n=0}^{d-1} \omega^{- (n^{2}-n) (d+1)/2}
\, |n\rangle \langle n|.
\end{equation}

Indeed this is the case, since one easily
verifies that
\begin{equation}
X = F^\dagger \, Z \, F,
\end{equation}
much in the spirit of the standard way of
looking at complementary variables in the
infinite-dimensional Hilbert space: the
position and momentum eigenstates are
Fourier transform one of the other.
On the other hand, the diagonal
transformation $V$ acts as a $Z$-right
shift:
\begin{equation}
\label{Vtrans}
X Z^{m} =  V^\dagger {}^{m} \, X \, V^{m} .
\end{equation}

The case $d=2$ needs minor modifications.
In fact, it turns out that one cannot find
a diagonal unitary transformation $V$ such that
$X \rightarrow X Z$. For this reason, instead
of $XY$ the matrix $Y$ is defined as $i X Z$,
so that $Y =  V^\dagger \, X \, V$, where $V$
is
\begin{equation}
V=
\left(
\begin{array}{rr}
1 & 0 \\
0 & -i
\end{array}
\right ) .
\end{equation}
The construction of multicomplementary
operators is otherwise identical to the
case where $d$ is an odd prime.

\section{Complementary operators
in composite dimensions}

\subsection{Constructing multicomplementary
operators}

For all its simplicity, the construction of
the previous Section fails if the dimension
of the system is a power of a prime. A simple
illustration of this is obtained in dimension
$4=2^{2}.$  According to equation~(\ref{CC}),
the operators $X$ and $Z$ are simply
\begin{equation}
X =
\left (
\begin{array}{cccc}
0 & 1 & 0 & 0 \\
0 & 0 & 1 & 0 \\
0 & 0 & 0 & 1 \\
1 & 0 & 0 & 0
\end{array}
\right ) ,
\qquad
Z =
\left(
\begin{array}{cccc}
1 & 0 & 0 & 0 \\
0 & i & 0 & 0 \\
0 & 0 & -1 & 0 \\
0 & 0 & 0 & -i
\end{array}
\right) .
\end{equation}
Then, for instance, $X^2 = - (X Z^{2})^{2}$
and  $( XZ^{3} )$ is proportional to $XZ,$ so
operators constructed following~(\ref{SCop})
no longer form disjoint sets. The root of
this failure can be traced to the fact
that $\mathbb{Z}_{4},$ the set of integers
modulo 4, does not form an algebraic field.
The same failure generally occurs for any
composite dimension $d= p^{n}$, where $p$
is a prime and $n$ is an integer. In short,
the construction of multicomplementary
operators cannot proceed by simply
taking powers of some basic elements.

However, we know there exists (up to
isomorphisms) exactly one field, written as
$\mathbb{F}_d$, with $d$ elements when $d = p^n$.
If $d=p$ is prime, the field essentially coincides
with $\mathbb{Z}_p$. We briefly recall the
minimum background needed to proceed.
For more details the reader is referred
to the pertinent literature~\cite{Lidl86}.
The field $\mathbb{F}_d$ can be
represented as the field of equivalence
classes of polynomials whose coefficients
belong to $\mathbb{Z}_p$. The product
in the multiplicative group $\mathbb{F}_d^{\ast}$
(i. e, excluding the zero) is defined as
the product of the corresponding polynomials
modulo a primitive polynomial of degree $n$
irreducible in $\mathbb{Z}_p$. In fact,
$\mathbb{F}_d^{\ast}$ is a cyclic
group of order $d-1$: it is generated by
powers of a primitive element $\alpha$,
which is a monic irreducible polynomial
of degree $n$. This establishes a natural
order for the field elements, and we use this
order to label the elements of a basis in
$\mathcal{H}_{d}$ as follows:
\begin{equation}
\label{mu_basis}
\{ |0 \rangle ,  | \alpha \rangle, | \alpha^2 \rangle,
\ldots, | \alpha^{d-1} \rangle \}  .
\end{equation}
Our solution to the problem of MUBs in composite
dimension consists in using  elements of $\mathbb{F}_{d}$,
instead of natural numbers, to label the classes
of complementary operators.

Next we define the trace of a field element
$\theta \in \mathbb{F}_d$ as
\begin{equation}
\tr (\theta ) = \theta + \theta^{p} + \theta^{p^2} +
\ldots + \theta^{p^{n-1}}.
\end{equation}
Note that we distinguish it from the trace of an
operator by the lower case "tr". The trace
has remarkably simple properties, the
most important for us being that it is linear
and that it is always an element of the prime
field $\mathbb{Z}_p$.

For the additive group in the field $\mathbb{F}_d$
we can introduce additive characters as a map
that fulfills
\begin{equation}
\chi ( \theta_1 ) \chi (\theta_2 ) =
\chi (\theta_1 + \theta_2 ) ,
\qquad
\theta_1, \theta_2 \in \mathbb{F}_d .
\end{equation}
All of these additive characters have the form
\begin{equation}
\label{defchi}
\chi (\theta )  = \exp \left[
\frac{2\pi i}{p} \tr ( \theta ) \right] .
\end{equation}
For future reference, we quote the property
\begin{equation}
\label{character_sum1}
\sum_{\theta \in \mathbb{F}_d}
\chi (\theta ) = 0 ,
\end{equation}
which leads to the relation
\begin{equation}
\label{character_sum2}
\sum_{k=0}^{d-2} \chi (\alpha^{k} \theta ) =
d \, \delta_{\theta,0}-1 .
\end{equation}

We start by introducing diagonal operators
with respect to the basis (\ref{mu_basis})
as follows:
\begin{equation}
Z_q  = |0\rangle \langle 0|
+ \sum_{k=1}^{d-1} \chi (\alpha^{q+k})
|\alpha^{k}\rangle \langle \alpha^{k}| ,
\qquad q = 0, \dots ,d-1 .
\end{equation}
This definition implies
\begin{equation}
Z_q |\alpha^{k} \rangle  =
\chi (\alpha^{q+k}) |\alpha^{k}\rangle ,
\end{equation}
and the combination property
\begin{equation}
Z_q Z_{q^\prime} \equiv
Z_{(q) + (q^{\prime })} = |0\rangle \langle 0|
+ \sum_{k=1}^{d-1}
\chi (\alpha^{q+k} + \alpha^{q^\prime + k})
|\alpha^{k} \rangle \langle \alpha^{k}| .
\end{equation}
These are quite natural generalizations
of the properties of matrices in the class
$\{ Z^k \}$ in~(\ref{SCop}), since the
$|\alpha^{k} \rangle$ are eigenstates of
$Z_q$.

In a similar fashion the operators $X_{q}$ are
defined as
\begin{equation}
\label{Xq}
X_q = \sum_{k=1}^{d-1}
|\alpha^{k} + \alpha^{q} \rangle
\langle \alpha^{k}| +
|\alpha^{q} \rangle \langle 0| ,
\qquad
q = 0,\ldots ,d-2,
\end{equation}
so that they act as operators shifting
$|\alpha^k \rangle$ to $|\alpha^{k} +
\alpha^{q} \rangle$ and satisfy the
combination rule
\begin{equation}
X_q X_{q^\prime} = X_{(q)+(q^\prime)} =
|\alpha^{q} + \alpha^{q^\prime} \rangle
\langle 0| + \sum_{k=1}^{d-1}
|\alpha^{k} + \alpha^{q} +
\alpha^{q^\prime} \rangle
\langle \alpha^{k}| .
\end{equation}
Because elements of the field close
under addition, $\alpha^{k} + \alpha^{q}$
is another element in the field: it must
be that there is some number $L(n)$,
called the Jacobi logarithm, such that
\begin{equation}
\alpha^{k} + \alpha^{q} =
\alpha^{k + L (q-k)} ,
\end{equation}
whenever $\alpha^{k} + \alpha^{q} \neq 0$.
In applications, $\alpha^{k}+\alpha^{q}$
can be found from the summation table of
$\mathbb{F}_d$.

The finite Fourier transform $F$, when expressed
in terms of the basis $|\alpha^{k}\rangle$, takes
the form
\begin{equation}
\label{F}
\fl
F = \frac{1}{\sqrt{d}} \left[
|0\rangle \langle 0| +
\sum_{k,k^\prime = 1}^{d-1}
\chi (\alpha^{k^\prime+k})
|\alpha^{k^\prime} \rangle \langle \alpha^{k}|
+ \sum_{k=1}^{d-1}
\left ( |0\rangle \langle \alpha^{k}| +
|\alpha^{k}\rangle \langle 0| \right )
\right] .
\end{equation}
This definition satisfies the natural property
\begin{equation}
F^{\dagger }F=FF^{\dagger }=I.
\end{equation}
The operator $F$ is set up to transform
the operators $Z_{q}$ into $X_{q}$:
\begin{equation}
X_q = F^\dagger \, Z_q \, F .
\end{equation}
The  Weyl form of the canonical commutation
relations is now
\begin{equation}
Z_q  X_{q^\prime} = \chi (\alpha^{q+q^\prime})
X_{q^\prime} Z_q .
\end{equation}
The operators $Z_{q}$ and $X_{q}$ have been
designed to be $d$-``periodic", in the sense that
\begin{equation}
Z_d = Z_0 ,
\qquad
X_d = X_0 .
\end{equation}

In fully analogy with the sets in  (\ref{SCop}),
we can generate operators from $X_{q}$ and $Z_{q}$;
they will be of the form $X_{q} Z_{r}$.
Linear independence and orthogonality
are guaranteed, in the sense that
[compare equation~(\ref{pairwise})]
\begin{eqnarray}
\label{DS}
\Tr ( Z_q Z_{q^\prime}^\dagger ) =
d \, \delta_{qq^\prime} ,
\qquad
\Tr ( X_q X_{q^\prime}^\dagger ) =
d \, \delta_{qq^\prime},
\nonumber  \\
& & \\
\Tr [ ( X_q  Z_r ) ( X_{q^\prime} Z_{r^\prime} ) ^\dagger ]
= d \, \delta_{q q^\prime} \delta_{r r^\prime}.
\nonumber
\end{eqnarray}
The commutation relations read
\begin{equation}
\label{CR}
[ X_q Z_r, X_{q^\prime} Z_{r^\prime} ] =
X_{(q)+(q^\prime)} Z_{(r)+(r^\prime)}
[\chi ( \alpha^{q^\prime + r}) -
\chi ( \alpha^{q + r^\prime} ) ] .
\end{equation}

It is clear from (\ref{DS}) and (\ref{CR}) that
the sets [compare equation~(\ref{SCop})]
\begin{eqnarray}
\label{CCF}
\{ Z_q \} ,
\qquad \qquad q = 0, \ldots, d-2, \nonumber \\
& & \\
\{ X_q Z_{q+r} \} ,
\qquad
q, r = 0, \ldots, d-2 , \nonumber
\end{eqnarray}
are disjoint and that every element of
a set with a fixed value  $r$ commutes
with every other element in the same set:
they define multicomplementary operators.

Finally, let us consider the form of the
diagonal operators similar to (\ref{Vtrans})
transforming $X_{q}$ to $X_{q}Z_{r}$. If we
restrict to odd dimensions, we have
\begin{equation}
\fl
V_q^{(r)} = |0 \rangle \langle 0| +
\sum_{k=1}^{d-1} \bar{\chi}( 2^{-1} \alpha^{r+2k-q})
|\alpha^{k} \rangle \langle \alpha^{k}| ,
\qquad q, r = 0,\ldots, d-2,
\label{Vq}
\end{equation}
where $\bar{\chi}$ means conjugate character
and $2^{-1}$ is an element of $\mathbb{Z}_p$;
in particular, if $p = 2N+1$ we have $2^{-1}= N+1$.
In this way, one can check that
\begin{equation}
V_{q+r}^{(q)}{}^\dagger X_q V_{q+r}^{(q)}  =
\chi (2^{-1} \alpha^{2q+r}) X_{q} Z_{q+r} .
\end{equation}

Using (\ref{F}) and (\ref{Vq}) we can generate
all the complementary bases. Indeed, if the vectors
$| \vec{\alpha} \rangle_{q}$ ($q = 0, \ldots, d-2$)
are the eigenstates of $Z_{0}$, then the whole set of
complementary bases can be obtained as follows:
\begin{equation}
V_{q}^{(0)} {}^\dagger \, F^\dagger
|\vec{\alpha} \rangle_q .  \label{odd_CB}
\end{equation}

\subsection{Complementary operators for two qubits}

We illustrate our approach with the simplest
case of a quantum system of composite dimension:
two qubits described in a four-dimensional
Hilbert space $\mathcal{H}_4$. To construct
multicomplementary operators, we start from
the field $\mathbb{F}_{4}$ containing
four elements. The polynomial
\begin{equation}
\label{pol2}
\theta^{2} + \theta + 1 = 0
\end{equation}
is irreducible in $\mathbb{Z}_2$ and the
primitive element $\alpha$ is defined as
a root of (\ref{pol2}).  In consequence
the four elements of $\mathbb{F}_{4}$
as in (\ref{mu_basis}) can be written as
\begin{equation}
\label{F4_mu}
\{0, 1, \alpha , \alpha+1 \} ,
\end{equation}
where we have taken into account
arithmetic modulo 2 and the fact
that if $\alpha$ satisfies equation~(\ref{pol2}),
then we have the relations
\begin{equation}
\alpha^2 = \alpha +1 ,
\qquad
\alpha^3 = 1 .
\end{equation}
A direct application of the definition
(\ref{defchi}) gives
\begin{equation}
\chi (0) =  1 ,
\quad
\chi (\alpha ) = - 1 ,
\quad
\chi (\alpha^2 ) = - 1 ,
\quad
\chi (\alpha^3 ) = 1 .
\end{equation}
Using a the representation  where
\begin{equation}
\label{rep}
\fl
|0 \rangle =
\left (
\begin{array}{c}
1 \\
0 \\
0 \\
0
\end{array}
\right) ,
\quad
|\alpha \rangle =
\left (
\begin{array}{c}
0 \\
1 \\
0 \\
0
\end{array}
\right) ,
\quad
|\alpha^2 \rangle =
\left (
\begin{array}{c}
0 \\
0 \\
1 \\
0
\end{array}
\right ) ,
\quad
| \alpha^3 \rangle =
\left (
\begin{array}{c}
0 \\
0 \\
0 \\
1
\end{array}
\right ) ,
\end{equation}
the matrices $Z_q$ are
\begin{eqnarray}
\label{Z4}
Z_0 =
\left (
\begin{array}{rrrr}
1 & 0 & 0 & 0 \\
0 & -1 & 0 & 0 \\
0 & 0 & -1 & 0 \\
0 & 0 & 0 & 1
\end{array}
\right ) ,
\qquad
Z_1 =
\left (
\begin{array}{rrrr}
1 & 0 & 0 & 0 \\
0 & -1 & 0 & 0 \\
0 & 0 & 1 & 0 \\
0 & 0 & 0 & -1
\end{array}
\right ) , \nonumber \\
 \\
Z_2 =
\left (
\begin{array}{rrrr}
1 & 0 & 0 & 0 \\
0 & 1 & 0 & 0 \\
0 & 0 & -1 & 0 \\
0 & 0 & 0 & -1
\end{array}
\right) .  \nonumber
\end{eqnarray}
The matrix realization of the Fourier
transform is
\begin{equation}
\label{F4_F}
F = \frac{1}{2}
\left (
\begin{array}{rrrrr}
1 & 1 & 1 & 1 \\
1 & -1 & 1 & -1 \\
1 & 1 & -1 & -1 \\
1 & -1 & -1 & 1
\end{array}
\right) ,
\end{equation}
and the matrices $X_q$ are
\begin{eqnarray}
\label{X4}
X_0 =
\left (
\begin{array}{rrrr}
0 & 0 & 0 & 1 \\
0 & 0 & 1 & 0 \\
0 & 1 & 0 & 0 \\
1 & 0 & 0 & 0
\end{array}
\right ) ,
\qquad
X_1 =
\left (
\begin{array}{rrrr}
0 & 1 & 0 & 0 \\
1 & 0 & 0 & 0 \\
0 & 0 & 0 & 1 \\
0 & 0 & 1 & 0
\end{array}
\right ) , \nonumber \\
& & \\
X_2 =
\left (
\begin{array}{rrrr}
0 & 0 & 1 & 0 \\
0 & 0 & 0 & 1 \\
1 & 0 & 0 & 0 \\
0 & 1 & 0 & 0
\end{array}
\right ) . \nonumber
\end{eqnarray}
The rest of the sets are routinely obtained
according (\ref{CCF}). For completeness
we quote all of them for this example:
\begin{eqnarray}
X_0 Z_0 =
\left (
\begin{array}{rrrr}
0 & 0 & 0 & 1 \\
0 & 0 & -1 & 0 \\
0 & -1 & 0 & 0 \\
1 & 0 & 0 & 0
\end{array}
\right ) ,
\qquad
X_1 Z_1 =
\left (
\begin{array}{rrrr}
0 & -1 & 0 & 0 \\
1 & 0 & 0 & 0 \\
0 & 0 & 0 & -1 \\
0 & 0 & 1 & 0
\end{array}
\right ) , \nonumber \\
X_2 Z_2 =
\left (
\begin{array}{rrrr}
0 & 0 & -1 & 0 \\
0 & 0 & 0 & -1 \\
1 & 0 & 0 & 0 \\
0 & 1 & 0 & 0
\end{array}
\right ) , \\
& & \nonumber \\
X_0 Z_1 =
\left (
\begin{array}{rrrr}
0 & 0 & 0 & -1 \\
0 & 0 & 1 & 0 \\
0 & -1 & 0 & 0 \\
1 & 0 & 0 & 0
\end{array}
\right ) ,
\qquad
X_1 Z_2 =
\left (
\begin{array}{rrrr}
0 & 1 & 0 & 0 \\
1 & 0 & 0 & 0 \\
0 & 0 & 0 & -1 \\
0 & 0 & -1 & 0
\end{array}
\right ) , \nonumber \\
X_2 Z_0 =
\left (
\begin{array}{rrrr}
0 & 0 & -1 & 0 \\
0 & 0 & 0 & 1 \\
1 & 0 & 0 & 0 \\
0 & -1 & 0 & 0
\end{array}
\right )  ,  \\
& & \nonumber \\
X_0 Z_2 =
\left (
\begin{array}{rrrr}
0 & 0 & 0 & -1 \\
0 & 0 & -1 & 0 \\
0 & 1 & 0 & 0 \\
1 & 0 & 0 & 0
\end{array}
\right ) ,
\qquad
X_1 Z_0 =
\left (
\begin{array}{rrrr}
0 & -1 & 0 & 0 \\
1 & 0 & 0 & 0 \\
0 & 0 & 0 & 1 \\
0 & 0 & -1 & 0
\end{array}
\right ) , \nonumber \\
X_2 Z_1 =
\left (
\begin{array}{rrrr}
0 & 0 & 1 & 0 \\
0 & 0 & 0 & -1 \\
1 & 0 & 0 & 0 \\
0 & -1 & 0 & 0
\end{array}
\right ) .
\end{eqnarray}

\subsection{Complementary operators for
three qubits}

Our next example is the case of three
qubits, the Hilbert space of which is
eight dimensional. For the field $\mathbb{F}_8$
the primitive element $\alpha$ is
a root of the following irreducible
polynomial on $\mathbb{F}_{2}$
\begin{equation}
\theta^{3} + \theta + 1 = 0 .
\end{equation}
In consequence, the elements of
$\mathbb{F}_{8}$ according to (\ref{SCop})
are
\begin{equation}
\label{F8_mu}
\{0, 1, \alpha , \alpha^2,  \alpha + 1,
\alpha^2 + \alpha , \alpha^2 + \alpha + 1,
\alpha^2 +1 \} ,
\end{equation}
where we have taken into account that
\begin{equation}
\fl
\alpha^3 = \alpha + 1,
\quad
\alpha^4 = \alpha^2 + \alpha ,
\quad
\alpha^5 = \alpha^2 + \alpha + 1,
\quad
\alpha^6 = \alpha^2 +1 ,
\quad
\alpha^7 = 1 .
\end{equation}
One can obtain again the additive characters
in a straightforward way
\begin{eqnarray}
\chi (0) = 1,
\quad
\chi ( \alpha ) = 1,
\quad
\chi ( \alpha^2 ) = 1,
\quad
\chi ( \alpha^3 ) = - 1 ,
\nonumber \\
\chi ( \alpha^4 ) = 1,
\quad
\chi ( \alpha^5 ) = - 1 ,
\quad
\chi ( \alpha^6 ) = - 1,
\quad
\chi ( \alpha^7 ) = - 1 .
\end{eqnarray}
Using the basis labeling as in (\ref{rep})
the matrix of the Fourier transform is
\begin{equation}
F=\frac{1}{\sqrt{2^3}}
\left (
\begin{array}{rrrrrrrr}
1 & 1 & 1 & 1 & 1 & 1 & 1 & 1 \\
1 & 1 & -1 & 1 & -1 & -1 & -1 & 1 \\
1 & -1 & 1 & -1 & -1 & -1 & 1 & 1 \\
1 & 1 & -1 & -1 & -1 & 1 & 1 & -1 \\
1 & -1 & -1 & -1 & 1 & 1 & -1 & 1 \\
1 & -1 & -1 & 1 & 1 & -1 & 1 & -1 \\
1 & -1 & 1 & 1 & -1 & 1 & -1 & -1 \\
1 & 1 & 1 & -1 & 1 & -1 & -1 & -1
\end{array}
\right ) .
\end{equation}
The matrices $Z_{q}$ ($q=0, \ldots,6$) are
\begin{eqnarray}
\fl
Z_0 = \mathrm{diag}(1,1,1,-1,1,-1,-1,-1),
\qquad
Z_1 = \mathrm{diag}(1,1,-1,1,-1,-1,-1,1),
\nonumber \\
\fl
Z_2 = \mathrm{diag}(1,-1,1,-1,-1,-1,1,1),
\qquad
Z_3 = \mathrm{diag}(1,1,-1,-1,-1,1,1,-1),
\nonumber \\
\fl
Z_4 = \mathrm{diag}(1,-1,-1,-1,1,1,-1,1),
\qquad
Z_{5} = \mathrm{diag}(1,-1,-1,1,1,-1,1,-1),
\nonumber \\
Z_{6} = \mathrm{diag}(1,-1,1,1,-1,1,-1,-1),
\end{eqnarray}
and, for instance,
\begin{equation}
X_0 =
\left (
\begin{array}{rrrrrrrr}
0 & 0 & 0 & 0 & 0 & 0 & 0 & 1 \\
0 & 0 & 0 & 1 & 0 & 0 & 0 & 0 \\
0 & 0 & 0 & 0 & 0 & 0 & 1 & 0 \\
0 & 1 & 0 & 0 & 0 & 0 & 0 & 0 \\
0 & 0 & 0 & 0 & 0 & 1 & 0 & 0 \\
0 & 0 & 0 & 0 & 1 & 0 & 0 & 0 \\
0 & 0 & 1 & 0 & 0 & 0 & 0 & 0 \\
1 & 0 & 0 & 0 & 0 & 0 & 0 & 0
\end{array}
\right ) .
\end{equation}
The rest of them can be easily worked
out.

\subsection{Complementary operators
for two qutrits}

Our final example is the case of two
qutrits, described in the nine-dimensional
Hilbert space $\mathcal{H}_9$. The primitive
element of the field $\mathbb{F}_9$
is a root of the following irreducible
polynomial on $\mathbb{F}_{3}$
\begin{equation}
\theta^2 + \theta + 2 = 0 ,
\end{equation}
so that the elements of $\mathbb{F}_9$ are
\begin{equation}
\label{F9_mu}
\{ 0, 1, 2, \alpha, 2 \alpha + 1,  2 \alpha + 2,
2, 2 \alpha , \alpha + 2, \alpha + 1,  \} ,
\end{equation}
and the additive characters are
\begin{eqnarray}
\chi (0) = 1,
\quad
\chi (\alpha ) = \bar{\omega},
\quad
\chi (\alpha^2 ) = 1,
\quad
\chi (\alpha^3 ) = \bar{\omega},
\quad
\chi ( \alpha^4 ) = \omega ,
\nonumber \\
\chi (\alpha^5) = \omega ,
\quad
\chi ( \alpha^6 ) = 1 ,
\quad
\chi (\alpha^7 ) = \omega ,
\quad
\chi ( \alpha^8 ) = \bar{\omega},
\end{eqnarray}
where $\omega = e^{2\pi i/3}$ and
the bar denotes complex conjugation.
Using the same basis as (\ref{rep}),
we have that the matrix
of the Fourier transform is
\begin{equation}
F=\frac{1}{3}
\left (
\begin{array}{rrrrrrrrr}
1 & 1 & 1 & 1 & 1 & 1 & 1 & 1 & 1 \\
1 & 1 & \bar{\omega} & \omega & \omega
& 1 & \omega & \bar{\omega} & \bar{\omega}\\
1 & \bar{\omega} & \omega & \omega & 1 &
\omega & \bar{\omega} & \bar{\omega} & 1 \\
1 & \omega & \omega & 1 & \omega & \bar{\omega} &
\bar{\omega} & 1 & \bar{\omega} \\
1 & \omega & 1 & \omega & \bar{\omega} &
\bar{\omega} & 1 & \bar{\omega} & \omega \\
1 & 1 & \omega & \bar{\omega} & \bar{\omega}
& 1 & \bar{\omega} & \omega & \omega \\
1 & \omega & \bar{\omega} & \bar{\omega} &
1 & \bar{\omega} & \omega & \omega & 1 \\
1 & \bar{\omega} & \bar{\omega} & 1 &
\bar{\omega} & \omega & \omega & 1 & \omega \\
1 & \bar{\omega} & 1 & \bar{\omega} & \omega
& \omega & 1 & \omega & \bar{\omega}
\end{array}
\right) .  \label{F9_F}
\end{equation}
The  operators $Z_{q}$ take the form
\begin{equation}
\fl
\begin{array}{ll}
Z_0 = \mathrm{diag} ( 1,\bar{\omega},1,
\bar{\omega},\omega ,\omega ,1,
\omega ,\bar{\omega}) ,
\qquad &
Z_1 = \mathrm{diag} ( 1,1,\bar{\omega},
\omega ,\omega ,1,\omega ,\bar{\omega},
\bar{\omega} ) , \\
Z_2 = \mathrm{diag} ( 1, \bar{\omega},
\omega ,\omega ,1, \omega ,\bar{\omega},
\bar{\omega},1 ) ,
\qquad &
Z_3 = \mathrm{diag} ( 1,\omega ,\omega ,
1,\omega ,\bar{\omega},\bar{\omega},1,
\bar{\omega} ) , \\
Z_4 =  \mathrm{diag} ( 1,\omega ,1,\omega ,
\bar{\omega},\bar{\omega},1, \bar{\omega},
\omega ) ,
\qquad &
Z_5 = \mathrm{diag} ( 1,1,\omega ,
\bar{\omega},\bar{\omega},1,\bar{\omega},
\omega ,\omega ) ,  \\
Z_6 = \mathrm{diag} ( 1,\omega ,\bar{\omega},
\bar{\omega},1,\bar{\omega},\omega ,
\omega ,1 ) ,
\qquad &
Z_7 = \mathrm{diag} ( 1,\bar{\omega},
\bar{\omega},1,\bar{\omega},\omega ,
\omega ,1,\omega ) .
\end{array}
\end{equation}
Note that $Z_{q}$ $(q = 1,\ldots,7)$ are obtained from
$Z_{0}$ by a cyclic permutation of the diagonal
elements, except the first element that always
remains 1. The $X_{q}$ are also easily constructed,
and we have, e. g.
\begin{equation}
X_{0}=\left(
\begin{array}{ccccccccc}
0 & 0 & 0 & 0 & 1 & 0 & 0 & 0 & 0 \\
0 & 0 & 0 & 0 & 0 & 0 & 1 & 0 & 0 \\
0 & 0 & 0 & 0 & 0 & 1 & 0 & 0 & 0 \\
0 & 0 & 1 & 0 & 0 & 0 & 0 & 0 & 0 \\
0 & 0 & 0 & 0 & 0 & 0 & 0 & 0 & 1 \\
0 & 0 & 0 & 1 & 0 & 0 & 0 & 0 & 0 \\
0 & 0 & 0 & 0 & 0 & 0 & 0 & 1 & 0 \\
0 & 1 & 0 & 0 & 0 & 0 & 0 & 0 & 0 \\
1 & 0 & 0 & 0 & 0 & 0 & 0 & 0 & 0
\end{array}
\right) .
\end{equation}
The diagonal operators $V_{q}$ have
the form
\begin{equation}
\fl
\begin{array}{ll}
V_0 = \mathrm{diag} ( 1, 1, \omega , 1,
\bar{\omega}, 1,\omega ,1,\bar{\omega} ) ,
\qquad &
V_1 = \mathrm{diag} ( 1, \bar{\omega},\omega ,
\omega ,\bar{\omega},\bar{\omega}, \omega ,
\omega , \bar{\omega} ) ,  \\
V_2 =  \mathrm{diag} ( 1,\omega ,1,
\bar{\omega},1,\omega ,1,\bar{\omega}, 1) ,
\qquad &
V_3 = \mathrm{diag} ( 1,\omega ,\omega ,
\bar{\omega},\bar{\omega},\omega,
\omega ,\bar{\omega},\bar{\omega} ) , \\
V_4 = \mathrm{diag} ( 1, 1, \bar{\omega}, 1,
\omega , 1, \bar{\omega}, 1, \omega ) ,
\qquad &
V_5 = \mathrm{diag} ( 1, \omega ,
\bar{\omega}, \bar{\omega}, \omega , \omega ,
\bar{\omega}, \bar{\omega}, \omega ) , \\
V_6 = \mathrm{diag} ( 1, \bar{\omega}, 1,
\omega , 1, \bar{\omega}, 1, \omega ,1 ) ,
\qquad &
V_7 = \mathrm{diag} ( 1, \bar{\omega},
\bar{\omega},\omega ,\omega ,\bar{\omega},
\bar{\omega},\omega ,\omega ) .
\end{array}
\end{equation}
Note that all the complementary bases can
be obtained directly from (\ref{odd_CB}); i. e.
applying $V_{q} F$ to the basis (\ref{F9_mu}).

\section{Multicomplemntary operators as tensor
products}

One can establish an isomorphism between
the form (\ref{CCF}) of complementary
operators and its representation in
terms of direct product of generalized
Pauli operators  (\ref{CC}). This isomorphism
can be put forward by showing a
one-to-one correspondence between
the basis (\ref{mu_basis}) and the
coefficients of the expansion of the
powers of the primitive element on,
for instance, the polynomial basis,
formed by $( 1, \alpha ,\alpha^{2}, \ldots,
\alpha^{n-1})$. In this way we get
\begin{equation}
\alpha^k  \mapsto
(c_0^{(k)}, c_1^{(k)}, \ldots,
c_{n-1}^{(k)}) ,
\qquad c_{l}^{(k)} \in \mathbb{Z}_{p},
\end{equation}
where
\begin{equation}
\alpha^{k} = \sum_{l=0}^{n-1} c_{l}^{(k)}
\alpha^{l}.
\end{equation}
This allows to rewrite the basis (\ref{mu_basis})
in the equivalent form
\begin{equation}
\{ |0\rangle , |c_0^{(k)}, c_1^{(k)}, \ldots,
c_{n-1}^{(k)}\rangle \} \equiv
\{ |0\rangle , |c_0^{(k)}\rangle |c_{1}^{(k)} \rangle
\ldots |c_{n-1}^{(k)}\rangle \},
 \label{c_basis}
\end{equation}
and the representation of (\ref{CCF}) as
a tensor product is now possible. This is due
to the fact that $\mathbb{F}_d$ is isomorphic
to $\mathbb{Z}_p \times \ldots \times \mathbb{Z}_p$,
with $n$ products.

It is worth noting that this isomorphism
can be settled also in other bases. For
example, the so-called normal basis,
obtained finding an element $\beta \in
\mathbb{F}_d$ such that
\begin{equation}
\{ \beta, \beta^p, \ldots, \beta^{p^{n -1}} \}
\end{equation}
is a basis of $\mathbb{F}_d$, is quite useful
in applications, since squaring a field element
can be easily accomplished by a right cyclic
shift. Of course, different bases lead
to different coefficients $c_{l}^{(k)}$,
different sets of commuting operators and
different factorizations of these operators
in tensor products.

\subsection{Two qubits}

To represent (\ref{Z4}) and (\ref{X4}) as
a tensor product we follow the described
above and write the basis (\ref{F4_mu}) in
the form (\ref{c_basis}). This yields
\begin{equation}
|0 \rangle = |00 \rangle ,
\quad
| \alpha \rangle = |01 \rangle ,
\quad
|\alpha^2 \rangle = |11\rangle ,
\quad
| \alpha^3 \rangle = |10\rangle .
\end{equation}
Then,
\begin{eqnarray}
Z_0 & \mapsto &  |00\rangle \langle 00|-
|01\rangle \langle 01| - |11 \rangle \langle 11|
+ |10 \rangle \langle 10| =
I \otimes \mathcal{Z}, \nonumber
\label{Z_c} \\
Z_1 & \mapsto & |00 \rangle \langle 00|-
|01 \rangle \langle 01| + |11\rangle \langle 11|
- |10 \rangle \langle 10| =
\mathcal{Z} \otimes \mathcal{Z}, \\
Z_{2} & \mapsto & |00 \rangle \langle 00|
+ |01 \rangle \langle 01|- |11 \rangle \langle 11|
-|10 \rangle \langle 10| =
\mathcal{Z} \otimes I, \nonumber
\end{eqnarray}
where $\mathcal{Z}= |0\rangle \langle 0|-
|1\rangle \langle 1|$. In a similar way
the representation of the $X_{k}$ is:
\begin{eqnarray}
\label{X_c}
X_0 & \mapsto & |11 \rangle \langle 01 | +
|01 \rangle \langle 11| + |00 \rangle
\langle 10| + |10 \rangle \langle 00| =
\mathcal{X} \otimes I,   \nonumber \\
X_1 & \mapsto & |00 \rangle \langle 01| +
|10 \rangle \langle 11| + |11 \rangle
\langle 10| + |01 \rangle \langle 00 | =
I\otimes \mathcal{X},   \\
X_2 & \mapsto & |10 \rangle \langle 01| +
|00 \rangle \langle 11| + |01 \rangle
\langle 10| + |11 \rangle \langle 00| =
\mathcal{X} \otimes \mathcal{X}, \nonumber
\end{eqnarray}
where $\mathcal{X} = |0 \rangle \langle 1|
+ |1\rangle \langle 0|$, so that $\mathcal{Z}
\mathcal{X} = -\mathcal{X} \mathcal{Z}$.

Note that such an asymmetrical correspondence
is a result of the Fourier transform (\ref{F4_F}),
which is not factorized into a direct product
of two Fourier operators.
The other complementary operators are obtained
as a simple product of (\ref{Z_c}) and (\ref{X_c}):
\begin{eqnarray}
( X_0 Z_0, X_1 Z_1, X_2 Z_2 ) \mapsto
( \mathcal{X} \otimes \mathcal{Z},
\mathcal{Z} \otimes \mathcal{Y},
\mathcal{Y} \otimes \mathcal{X} ) ,
\nonumber \\
( X_0 Z_1, X_1 Z_2, X_2 Z_0 ) \mapsto
( \mathcal{Y} \otimes \mathcal{Z},
\mathcal{Z} \otimes \mathcal{X},
\mathcal{X} \otimes \mathcal{Y} ) , \\
( X_0 Z_2, X_1 Z_0, X_2 Z_1 ) \mapsto
( \mathcal{Y}\otimes I,
I\otimes \mathcal{Y},
\mathcal{Y}\otimes \mathcal{Y} ) , \nonumber
\end{eqnarray}
where $\mathcal{Y}=\mathcal{X}\mathcal{Z}$.

We complete this example using the normal
basis $\{ \beta, \beta^2 \}$, with
$\alpha =  \beta, \alpha^2 = \beta^2,
\alpha^3 = \beta + \beta^2$. Then
\begin{equation}
|0 \rangle = |00 \rangle ,
\quad
| \alpha \rangle = |10 \rangle ,
\quad
|\alpha^2 \rangle = |01\rangle ,
\quad
| \alpha^3 \rangle = |11 \rangle ,
\end{equation}
and the new tensor product representatives are
\begin{equation}
\label{norm}
\begin{array}{ll}
Z_0 \mapsto  \mathcal{Z} \otimes \mathcal{Z},
\qquad \qquad &
X_0 \mapsto  \mathcal{X} \otimes \mathcal{X}, \\
Z_1 \mapsto \mathcal{Z} \otimes I ,
\qquad \qquad &
X_1 \mapsto \mathcal{X} \otimes I , \\
Z_2 \mapsto  I \otimes \mathcal{Z} ,
\qquad \qquad &
X_2 \mapsto I \otimes \mathcal{X}  ,
\end{array}
\end{equation}
while the rest of complementary operators
are given by
\begin{eqnarray}
( X_0 Z_0, X_1 Z_1, X_2 Z_2 ) \mapsto
( \mathcal{Y} \otimes \mathcal{Y},
\mathcal{Y} \otimes I,
I \otimes \mathcal{Y} ) ,
\nonumber \\
( X_0 Z_1, X_1 Z_2, X_2 Z_0 ) \mapsto
( \mathcal{Y} \otimes \mathcal{Z},
\mathcal{Z} \otimes \mathcal{X},
\mathcal{X} \otimes \mathcal{Y} ) , \\
( X_0 Z_2, X_1 Z_0, X_2 Z_1 ) \mapsto
( \mathcal{Z} \otimes \mathcal{X} ,
\mathcal{Y} \otimes \mathcal{X},
\mathcal{X} \otimes \mathcal{Z} ) . \nonumber
\end{eqnarray}
The symmetrical aspect of (\ref{norm})
is a consequence of the factorization of
Fourier transform in this basis. In fact,
we have
\begin{equation}
F = F_2 \otimes F_2 ,
\end{equation}
with
\begin{equation}
F_2 = \frac{1}{\sqrt{2}} (| 0 \rangle \langle 0 |
+ | 0 \rangle \langle 1 |+ | 1 \rangle \langle 0 |
- | 1 \rangle \langle 1 |).
\end{equation}

\subsection{Three qubits}

Mapping to tensor product form  is obtained
for this case in a manner similar to
$\mathbb{F}_{4}$, by establishing a correspondence
between the bases (\ref{mu_basis}) and (\ref{c_basis}).
Taking into account (\ref{F8_mu})
we obtain (by expanding over the
polynomial basis $\{1, \alpha , \alpha^2 \}$):
\begin{eqnarray}
|0 \rangle = |000 \rangle ,
\quad
| \alpha \rangle = |010 \rangle ,
\quad
| \alpha^2 \rangle = |001\rangle ,
\quad
| \alpha^3 \rangle = |110\rangle ,
\nonumber \\
| \alpha^4 \rangle = |011\rangle ,
\quad
| \alpha^5 \rangle = |111\rangle ,
\quad
| \alpha^6 \rangle = |101 \rangle ,
\quad
| \alpha^7 \rangle = |100 \rangle .
\end{eqnarray}
In this way one gets
\begin{equation}
\begin{array}{ll}
Z_0 \mapsto  \mathcal{Z} \otimes I
\otimes I ,
\qquad \qquad &
X_0 \mapsto \mathcal{X} \otimes I
\otimes I ; \\
Z_1 \mapsto I \otimes I \otimes
\mathcal{Z} ,
\qquad \qquad &
X_1 \mapsto I \otimes \mathcal{X}
\otimes I ; \\
Z_2 \mapsto I \otimes \mathcal{Z}
\otimes I ,
\qquad \qquad &
X_2 \mapsto I \otimes I
\otimes \mathcal{X}; \\
Z_3 \mapsto  \mathcal{Z} \otimes I
\otimes \mathcal{Z} ,
\qquad \qquad &
X_3 \mapsto \mathcal{X} \otimes
\mathcal{X} \otimes I ; \\
Z_4 \mapsto I \otimes\mathcal{ Z}
\otimes \mathcal{Z},
\qquad \qquad &
X_4 \mapsto  I \otimes \mathcal{X}
\otimes \mathcal{X} ; \\
Z_5 \mapsto  \mathcal{Z} \otimes
\mathcal{Z} \otimes \mathcal{Z} ,
\qquad \qquad &
X_5 \mapsto\mathcal{X} \otimes
\mathcal{X} \otimes \mathcal{X}; \\
Z_6 \mapsto  \mathcal{Z} \otimes
\mathcal{Z} \otimes I ,
\qquad \qquad &
X_6 \mapsto\mathcal{X} \otimes I
\otimes \mathcal{X},
\end{array}
\end{equation}
where the operators $\mathcal{Z}$ and
$\mathcal{X}$ are those of the two-qubit
example.

The other sets of commutative operators
are as follows:
\begin{equation}
\fl
\begin{array}{lll}
\{ X_q Z_q \} & \mapsto &
( \mathcal{Y} \, I  \, I,
I \, \mathcal{X} \, \mathcal{Z},
I \, \mathcal{Z} \, \mathcal{X},
\mathcal{Y} \, \mathcal{X} \, \mathcal{Z},
I \, \mathcal{Y} \, \mathcal{Y},
\mathcal{Y} \, \mathcal{Y} \,\mathcal{Y},
\mathcal{Y} \, \mathcal{Z} \,\mathcal{X}) ,\\
\{X_{q}Z_{q+1} \} & \mapsto &
( \mathcal{X} \, I \, \mathcal{Z},
I \,\mathcal{Y} \, I,
\mathcal{Z} \,I \,\mathcal{Y},
\mathcal{X} \, \mathcal{Y} \, \mathcal{Z},
\mathcal{Z} \, \mathcal{Y} \, \mathcal{Y},
\mathcal{Y} \, \mathcal{Y} \, \mathcal{X},
\mathcal{Y} \, I \, \mathcal{X} ) ,  \\
\{X_q Z_{q+2} \} & \mapsto &
( \mathcal{X} \,\mathcal{Z} \, I,
\mathcal{Z} \, \mathcal{X} \, \mathcal{Z},
I \, \mathcal{Z} \, \mathcal{Y},
\mathcal{Y} \, \mathcal{Y} \, \mathcal{Z},
\mathcal{Z} \, \mathcal{Y} \, \mathcal{X},
\mathcal{Y} \, \mathcal{X} \, \mathcal{X},
\mathcal{X} \, I \, \mathcal{Y} ) , \\
\{X_q Z_{q+3} \} & \mapsto &
( \mathcal{Y} \, I \, \mathcal{Z},
I \, \mathcal{Y} \, \mathcal{Z},
\mathcal{Z} \, \mathcal{Z} \, \mathcal{Y},
\mathcal{Y} \, \mathcal{Y} \, I,
\mathcal{Z} \, \mathcal{X} \, \mathcal{X},
\mathcal{X} \, \mathcal{X} \, \mathcal{Y},
\mathcal{X} \, \mathcal{Z} \, \mathcal{X} ) , \\
\{ X_q Z_{q+4} \} & \mapsto &
( \mathcal{X} \, \mathcal{Z} \, \mathcal{Z},
\mathcal{Z} \, \mathcal{Y} \, \mathcal{Z},
\mathcal{Z} \, \mathcal{Z} \, \mathcal{X},
\mathcal{Y} \, \mathcal{X} \,I,
I \, \mathcal{X} \,\mathcal{Y},
\mathcal{X} \, \mathcal{Y} \, \mathcal{X},
\mathcal{Y} \, I \, \mathcal{Y} ) , \\
\{X_q Z_{q+5} \} & \mapsto &
( \mathcal{Y} \,\mathcal{Z} \, \mathcal{Z},
\mathcal{Z} \, \mathcal{Y} \,I,
\mathcal{Z} \,I \,\mathcal{X},
\mathcal{X} \, \mathcal{X} \, \mathcal{Z},
I \, \mathcal{Y} \,\mathcal{X},
\mathcal{Y} \, \mathcal{X} \, \mathcal{Y},
\mathcal{X} \, \mathcal{Z} \, \mathcal{Y} ) , \\
\{X_q Z_{q+6} \} & \mapsto &
( \mathcal{Y} \, \mathcal{Z} \, I,
\mathcal{Z} \, \mathcal{X} \,I,
I \, I \, \mathcal{Y},
\mathcal{X} \, \mathcal{Y} \,I ,
\mathcal{Z} \, \mathcal{X} \, \mathcal{Y},
\mathcal{X} \, \mathcal{Y} \, \mathcal{Y},
\mathcal{Y} \, \mathcal{Z} \, \mathcal{Y} ) ,
\end{array}
\end{equation}
where $q = 0, \ldots, 6$ and we have omitted
the symbol $\otimes$ to simplify the writing.
Again the Fourier transform is not factorized
in this basis, but it is possible to find a
different basis where $F$ factorizes.

\subsection{Two qutrits}

The representation of $Z_{q}$ and $X_{q}$
operators in terms of tensor product is
obtained using the expansion in
the polynomial basis $\{1, \alpha \}$:
\begin{eqnarray}
|0 \rangle = |00 \rangle ,
\quad
| \alpha \rangle = | 01 \rangle ,
\quad
|\alpha^2 \rangle = | 12 \rangle ,
\quad
| \alpha^3 \rangle = | 22 \rangle ,
\quad
| \alpha^4 \rangle = | 20 \rangle ,
\nonumber \\
| \alpha^5 \rangle = | 02 \rangle ,
\quad
| \alpha^6 \rangle = | 21 \rangle ,
\quad
| \alpha^7 \rangle = | 11 \rangle ,
\quad
| \alpha^8 \rangle = | 10 \rangle ,
\end{eqnarray}
which leads to
\begin{equation}
\begin{array}{ll}
Z_0 \mapsto \mathcal{Z}^2 \, \mathcal{Z}^2;
\qquad \qquad &
X_0 \mapsto \mathcal{X} \, I; \\
Z_1 \mapsto \mathcal{Z}^2 \, I,
\qquad \qquad &
X_1 \mapsto I \, \mathcal{X}; \\
Z_2 \mapsto I \, \mathcal{Z}^2 ,
\qquad \qquad &
X_2 \mapsto \mathcal{X} \, \mathcal{X}^2 ; \\
Z_3 \mapsto \mathcal{Z}^2 \, \mathcal{Z},
\qquad \qquad &
X_3 \mapsto \mathcal{X}^2 \, \mathcal{X}^2 , \\
Z_4 \mapsto \mathcal{Z} \, \mathcal{Z} ,
\qquad \qquad &
X_4 \mapsto \mathcal{X}^2 \, I , \\
Z_5 \mapsto \mathcal{Z} \, I ,
\qquad \qquad &
X_5 \mapsto I \, \mathcal{X}^2 ; \\
Z_6 \mapsto I \, \mathcal{Z} ,
\qquad \qquad &
X_6 \mapsto \mathcal{X}^2 \, \mathcal{X} , \\
Z_7 \mapsto \mathcal{Z} \, \mathcal{Z}^2,
\qquad \qquad &
X_7 \mapsto \mathcal{X} \, \mathcal{X},
\end{array}
\end{equation}
where now
\begin{eqnarray}
\mathcal{Z} = |0 \rangle \langle 0| +
\omega | 1 \rangle \langle 1 | +
\bar{\omega} | 2 \rangle \langle 2|,
\nonumber \\
\\
\mathcal{X} = | 1 \rangle \langle 0 |
+ | 2 \rangle \langle 1 |
+ | 0 \rangle \langle 2|, \nonumber
\end{eqnarray}
so (\ref{CC}) holds, since $\mathcal{Z}\mathcal{X}
=\omega \mathcal{X}\mathcal{Z}.$

The sets of commutative operators are
\begin{equation}
\fl
\begin{array}{rll}
\{ X_q Z_q \} & \mapsto  &
( \mathcal{W} \, \mathcal{Z}^2,
\mathcal{Z}^2 \, \mathcal{X},
\mathcal{X} \, \mathcal{Y}^2,
\mathcal{Y}^2 \, \mathcal{W}^2,
\mathcal{W}^2 \, \mathcal{Z},
\mathcal{Z} \, \mathcal{X}^2,
\mathcal{X}^2 \, \mathcal{Y},
\mathcal{Y}\, \mathcal{W} ) , \\
\{ X_q Z_{q+1} \} & \mapsto &
( \mathcal{W} \, I,I\,
\mathcal{W},
\mathcal{W}\, \mathcal{W}^{2},
\mathcal{W}^{2} \, \mathcal{W}^{2},
\mathcal{W}^{2} \, I,
I \, \mathcal{W}^{2},
\mathcal{W}^{2} \, \mathcal{W},
\mathcal{W} \, \mathcal{W} ) , \\
\{ X_q Z_{q+2} \} & \mapsto &
( \mathcal{X} \, \mathcal{Z}^2,
\mathcal{Z}^2 \, \mathcal{Y},
\mathcal{Y} \, \mathcal{W}^2,
\mathcal{W}^2 \, \mathcal{X}^2,
\mathcal{X}^2 \, \mathcal{Z},
\mathcal{Z} \, \mathcal{Y}^2,
\mathcal{Y}^2 \, \mathcal{W},
\mathcal{W} \, \mathcal{X} ) , \\
\{ X_q \mathcal{Z}_{q+3} \} & \mapsto &
( \mathcal{W} \, \mathcal{Z},
\mathcal{Z} \, \mathcal{Y},
\mathcal{Y} \, \mathcal{X}^2,
\mathcal{X}^2 \, \mathcal{W}^2,
\mathcal{W}^2 \, \mathcal{Z}^2,
\mathcal{Z}^2 \, \mathcal{Y}^2,
\mathcal{Y}^2 \, \mathcal{X},
\mathcal{X} \, \mathcal{W} ) , \\
\{ X_q Z_{q+4} \} & \mapsto &
( \mathcal{Y} \, \mathcal{Z},
\mathcal{Z} \, \mathcal{X},
\mathcal{X} \, \mathcal{W}^2,
\mathcal{W}^2 \, \mathcal{Y}^2,
\mathcal{Y}^2 \, \mathcal{Z}^2,
\mathcal{Z}^2 \, \mathcal{X}^2,
\mathcal{X}^2 \, \mathcal{W},
\mathcal{W} \, \mathcal{Y} ) , \\
\{ X_q Z_{q+5} \} & \mapsto &
( \mathcal{Y} \, I,
I \, \mathcal{Y},
\mathcal{Y} \, \mathcal{Y}^2,
\mathcal{Y}^2 \, \mathcal{Y}^2,
\mathcal{Y}^2 \, I,
I \, \mathcal{Y}^2,
\mathcal{Y}^2 \, \mathcal{Y},
\mathcal{Y} \, \mathcal{Y} ) , \\
\{ X_q Z_{q+6} \} & \mapsto &
( \mathcal{X} \, \mathcal{Z},
\mathcal{Z} \, \mathcal{W},
\mathcal{W} \, \mathcal{Y}^2,
\mathcal{Y}^2 \, \mathcal{X}^2,
\mathcal{X}^2 \, \mathcal{Z}^2,
\mathcal{Z}^2 \, \mathcal{W}^2,
\mathcal{W}^2 \, \mathcal{Y},
\mathcal{Y} \, \mathcal{X} ) , \\
\{ X_q Z_{q+7}  \} & \mapsto &
( \mathcal{Y} \, \mathcal{Z}^2,
\mathcal{Z}^2 \, \mathcal{W},
\mathcal{W} \, \mathcal{X}^2,
\mathcal{X}^2 \, \mathcal{Y}^2,
\mathcal{Y}^2 \, \mathcal{Z},
\mathcal{Z} \, \mathcal{W}^2,
\mathcal{W}^2 \, \mathcal{X} ) ,
\end{array}
\end{equation}
where $q = 0, \ldots, 7$, $\mathcal{Y}= \mathcal{X}
\mathcal{Z}$, $\mathcal{W} = \mathcal{X} \mathcal{Z}^2$
and we have omitted global phases that appear in the
product of operators. Note that the above sets of
commuting operators are different from those listed
in~\cite{Lawrence04}.

\section{Concluding remarks}

In this paper we have solved the problem of
existence and construction of sets of MUBs
in composite dimensions.  Inspired by the
the approach developed in~\cite{Ban02}, in
which an interesting explicit construction
was shown for prime dimension, we have
used algebraic field extensions to produce
a solution for composite dimensions.
Although other constructive algorithms
for solving the MUBs problem in
composite dimension have appeared,
the one presented here does not resort
to dual basis, and so is completely
analogous to the prime-dimensional case.
We have also provided a simple scheme
to cast the sets of MUBs observables
as tensor products of generalized Pauli
matrices. This tensor products can
be quite different when we represent
the finite field in different bases.

Another major advantage of our approach relies
on the use of the finite Fourier transform
and a diagonal shift operator as  maps
between maximal classes of commuting operators.
This is in full agreement with our understanding
of complementarity in the infinite-dimensional
case.

Of course, there are still open questions:
the non-prime dimensional case is a
challenging issue, or what happens in
the limit of high dimensions. In any
case, the picture developed in this paper
is a valuable tool to deal with concepts
such as entanglement or separability
in finite-dimensional systems.

\ack
The authors have benefited from stimulating
discussions with Prof. Gunnar Bj\"ork.
The work of Hubert de Guise is supported by
NSERC of Canada.

\end{document}